\documentclass[aps,twocolumn,superscriptaddress,showpacs,amsmath,footinbib]{revtex4}

\usepackage{epsf,latexsym}
\usepackage{amssymb,amsmath}
\usepackage{times}

\newcommand{\bra}[1]{\langle #1 |}
\newcommand{\ket}[1]{| #1 \rangle}

\newcommand\h{{\cal H}}

\newcommand\s{{\cal S}}
\newcommand\A{{\cal A}}

\newcommand\diag{{\mbox{diag\,}}}

\newcommand{\ignore}[1]{}

\newcommand{\ra}{{\rightarrow}}

\newcommand{\be}{\begin{equation}}
\newcommand{\ee}{\end{equation}}
\newcommand{\ba}{\begin{eqnarray}}
\newcommand{\ea}{\end{eqnarray}}

\newtheorem{proposition}{Proposition}

\newtheorem{corollary}{Corollary}

\def\CC{{\rm\kern.24em \vrule width.04em height1.46ex depth-.07ex
    \kern-.30em C}}
\def\P{{\rm I\kern-.25em P}}

\def\RR{{\rm
         \vrule width.04em height1.58ex depth-.0ex
         \kern-.04em R}}

\def\bbbc{{\mathchoice {\setbox0=\hbox{$\displaystyle\rm C$}\hbox{\hbox
to0pt{\kern0.4\wd0\vrule height0.9\ht0\hss}\box0}}
{\setbox0=\hbox{$\textstyle\rm C$}\hbox{\hbox
to0pt{\kern0.4\wd0\vrule height0.9\ht0\hss}\box0}}
{\setbox0=\hbox{$\scriptstyle\rm C$}\hbox{\hbox
to0pt{\kern0.4\wd0\vrule height0.9\ht0\hss}\box0}}
{\setbox0=\hbox{$\scriptscriptstyle\rm C$}\hbox{\hbox
to0pt{\kern0.4\wd0\vrule height0.9\ht0\hss}\box0}}}}

\def\bbbz{{\mathchoice {\hbox{$\sf\textstyle Z\kern-0.4em Z$}}
{\hbox{$\sf\textstyle Z\kern-0.4em Z$}}
{\hbox{$\sf\scriptstyle Z\kern-0.3em Z$}}
{\hbox{$\sf\scriptscriptstyle Z\kern-0.2em Z$}}}}

\newcommand{\putfig}[2]{$$\leavevmode\hbox{\epsfxsize=#2 cm
   \epsffile{#1.eps}}$$}
\newcommand{\insertfig}[2]{\leavevmode \vcenter{\hbox{\epsfxsize=#2 cm
   \epsffile{#1.eps}}}}

\begin{document}

\title{Encoding graphs into quantum states: an axiomatic approach}

\author{Radu Ionicioiu}
\affiliation{Institute for Quantum Computing, University of Waterloo, Waterloo, Ontario N2L 3G1, Canada}
\affiliation{Department of Applied Mathematics, University of Waterloo, Waterloo, Ontario N2L 3G1, Canada}

\author{Tim P.~Spiller}
\affiliation{School of Physics and Astronomy, University of Leeds, Leeds LS2 9JT, UK}

\begin{abstract}
A fundamental problem in quantum information is to describe efficiently multipartite quantum states. An efficient representation in terms of graphs exists for several families of quantum states (graph, cluster, stabilizer states), motivating us to extend this construction to other classes. We introduce an axiomatic framework for mapping graphs to quantum states of a suitable physical system. Starting from three general axioms we derived a rich structure which includes and generalizes several classes of multipartite entangled state, like graph/stabilizer states, Gaussian cluster states, quantum random networks and projected entangled pair states (PEPS). Due to its flexibility we can extend the present formalism to include directed and weighted graphs.
\end{abstract}


\maketitle

\section{Introduction}

A key feature of quantum systems is the exponentially large Hilbert space required to describe them. This property is a mixed blessing. On the one hand, quantum systems can efficiently simulate quantum dynamics with polynomial resources. On the other, characterization of an arbitrary state of the Hilbert space needs an exponentially large number of parameters, a well-known problem in quantum tomography. Thus it becomes crucial to characterize efficiently quantum states. One such class of $n$-qubit states are the {\em stabilizer states}, which are completely (and efficiently) characterized by $n$ commuting stabilizer operators \cite{stabil}. Stabilizer states are local equivalent to another important family, namely {\em graph states} \cite{graph1, graph2, graph3, 1wqc}. This implies that one can graphically represent stabilizer states as graph states (up to local unitaries), giving them a visual impact: by merely looking at the graph one can glimpse the entanglement structure of the state (which is not easy to capture from writing the state in the computational basis, for example). Although graph states are successful in describing a large family of entangled states (cluster states, GHZ, CSS codes \cite{css}) they fail to include several important quantum states, like $W_n$ \cite{w1,w2} or Dicke states \cite{dicke1,dicke2}. It is thus natural to explore to what extent we can go beyond graph states and define a new class of entangled states, while at the same time still keeping the appealing connection to graphs.

A second motivation comes from graph theory. There are several hard problems in graph theory (graph isomorphism, 3-colorability) for which an efficient (polynomial) solution would be highly desirable. Several authors attempted to solve problems like graph isomorphism using a physically motivated approach \cite{tr, tr2, gudkov, shiau}. The intuition is to encode a graph into the quantum state of a system (bosons hopping on a graph \cite{shiau}, quantum walks \cite{q_walk1, q_walk2, q_walk3}) and to derive properties of the underlying graph $G$ from the associated quantum state $\ket{G}$.

\begin{figure}
\putfig{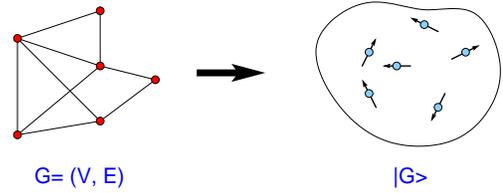}{6.5}
\caption{(color online). Given a graph $G= (V,E)$, how do we map it to a quantum state $\ket{G}\in \h$, for a suitable quantum system described by $\h$?}
\label{g2s}
\end{figure}

These considerations bring us to the problem we investigate in this article (Fig.\ref{g2s}): {\em Given a graph $G$, how do we associate to it a state $\ket{G}$ of a suitable quantum system?} Our approach to address this question is axiomatic. Starting from a few general principles we construct an axiomatic framework for mapping (encoding) graphs to states of a quantum system.

The structure of the article is the following. After a quick overview of graph theory and the main notations, in Section \ref{s2} we develop the framework starting from three intuitive axioms. In Section \ref{s3} we show that a number of well known quantum states, like graph states, Gaussian and continuous-variable cluster states, quantum random networks, projected entangled pair states (PEPS), share the structure introduced here and emerge naturally from our construction. In Section \ref{s4} we extend this approach to directed and weighted graphs and finally we conclude in Section \ref{s5}.

\noindent{\em Notations and background.} A graph is a pair $G= (V,E)$, where $V (E)$ is the set of vertices (edges). The {\em order} $|G|$ of the graph is the number of its vertices. The empty graph $E_n= (V, \varnothing)$ has $n$ vertices and no edges. $K_n$ is the complete graph with $n$ vertices, e.g., $K_2$ is an edge and $K_3$ a triangle. In a {\em regular graph} each vertex has the same {\em degree}, i.e., number of incident edges. In a {\em directed graph} the edges have an orientation, hence $(x,y)\in E$ is an ordered pair. The edges in a {\em weighted graph} have an associated label (weight).

A graph $G$ is described by its adjacency matrix $\A(G)$, defined as $\A(G)_{ij}= 1$, if the edge $(i,j)\in E$ and $\A(G)_{ij}= 0$ otherwise. For {\em simple graphs} (undirected graphs without loops and at most one edge between any pair of vertices) $\A$ is symmetric with zero diagonal. Two graphs of order $n$ are {\em isomorphic} $G\simeq G'$, iff there is a vertex permutation $P\in \s_n$ (the symmetric group of $n$ objects) preserving edge incidence: $\A(G')= P \A(G) P^{-1}$. A graph {\em automorphism} is an isomorphism of $G$ into itself, $[ \A(G), P ] =0$, i.e., a graph symmetry.

\section{Axioms}\label{s2}

In this section we develop the main framework for mapping graphs to quantum states. Starting from three intuitive principles (or axioms), we derive a structure which is general enough to include several well-known families of multipartite quantum states. In Section \ref{s4} we will see that this structure is also flexible enough to construct novel states starting from directed and weighted graphs.

We now formulate the problem more precisely. Given a graph $G$, we want to: (i) find a suitable Hilbert space $\h$, and (ii) associate to $G$ a state $\ket{G}\in \h$. For simplicity we consider only pure states $\rho(G)= \ket{G}\bra{G}$ (a mixed state can be obtained from the pure state of a larger system by a suitable partial trace).

For point (i), a preliminary question is the following: given a graph $G$, how large should the associated Hilbert space be? An answer requires an estimate of the size of the graph space. For a simple graph of order $n$ the number of edges satisfies $0 \le L \le n(n-1)/2$. Thus the number of simple graphs of order $n$ is ${\cal N}(n)\le 2^{n(n-1)/2}$; this is an upper bound and does not take into account graph isomorphisms. Using a similar argument, the number of graphs with $n$ vertices and $L$ edges is bounded by ${\cal N}(n,L) \le {n(n-1)/2 \choose L}$. Thus, if we want to capture enough of the structure of the graph space, the Hilbert space of the quantum system should have a similar dimension.

The second point (ii) is more difficult. {\em A priori} there are several possible ways to tackle the problem \cite{graph1, graph2, graph3, kitaev, severini1, severini2, severini3, tr, tr2, ggz}. As it is not clear which method is best suited (mapping vertices to qubits? edges to qubits? other?), we reframe the problem by asking a different question: {\em What properties should the mapping $G\ra \ket{G}$ have?} This gives a new perspective and provides us with a starting point. In the following we discuss three desirable properties for our mapping, quantified as a set of axioms.

Suppose we have two disjoint graphs $G_1$ and $G_2$. What is a natural way to map the disjoint sum of two graphs $G_1 \uplus G_2$ to a quantum state $\ket{G_1 \uplus G_2}$? As physicists we are used to think in terms of subsystems and the fourth postulate of quantum mechanics \cite{nc} gives us a hint in the right direction. Thus we choose as the first axiom the following:

\begin{quote}
{\bf A1:} {\bf Separability (tensor product).} For a disjoint sum of two graphs $G_1, G_2$ we have:
\be
\ket{G_1 \uplus G_2}= \ket{G_1}\otimes \ket{G_2}
\label{a1}
\ee
\end{quote}

This property immediately implies the following corollary:
\begin{corollary}
If $E_n= (V, \varnothing)$ is the empty graph on $n$ vertices, then
\[ \ket{E_n}= \ket{\psi_1}\otimes \ldots \otimes \ket{\psi_n}\]
\end{corollary}

Since the empty graph $E_n$ is mapped to a product state, this settles the initial question in favor of the mapping {\em vertices $\mapsto$ qudits}, as each vertex $i$ has an associated quantum state $\ket{\psi_i}$. Hence we have:
\begin{corollary}
Given a graph $G =(V, E)$ of order $n$, we associate to each vertex $i \in V$ a Hilbert space $\h_i$. The total Hilbert space is
\be
\h= \bigotimes_{i=1}^n \h_i
\ee
\end{corollary}
It is worth stressing the power of the separability axiom {\bf A1} -- this axiom alone automatically implies the tensor product structure of the Hilbert space.

While the first axiom is inspired by quantum theory, the second one has its roots in graph theory. Graph isomorphism is a central concept and we would like to preserve it under the mapping. The second axiom naturally captures this property:
\begin{quote}
{\bf A2:} {\bf Graph isomorphism.} If $G_1\simeq G_2$ are isomorphic, the corresponding density operators $\rho_{1,2}= \ket{G_{1,2}} \bra{G_{1,2}}$ satisfy:
\be
\rho_2= D(P) \rho_1 D(P)^{-1}
\label{a2}
\ee
where $D(P)$ is a matrix representation of the permutation $P \in \s_n$ mapping $G_1$ to $G_2$, i.e., $\A(G_2)= P \A(G_1) P^{-1}$.
\end{quote}
Notice that following Corollary 2 the permutation $P$ is well defined, as it interchanges the Hilbert spaces of the corresponding vertices. Also, expressing {\bf A2} in terms of  the density matrix $\rho$ naturally avoids possible extra phases related to the statistics of identical particles. 

Axiom {\bf A2} implies straightforwardly the following corollary for graph automorphisms:
\begin{corollary}
If $P\in \s_n$ is a graph automorphism of $G$, then we have
\be
[\rho, D(P)]= 0
\label{d2a}
\ee
\end{corollary}
Graph symmetries are thus captured naturally -- as expected, they commute with $\rho$.

Corollary 2 alone does not impose any restriction on the Hilbert spaces $\h_i$ (they are completely arbitrary). However, together with Corollary 3 (graph automorphism) it implies that all the Hilbert spaces are identical, $\h_i= \h_1, \forall i$. This follows from the automorphism group of the empty graph $E_n$ which is the whole symmetric group ${\cal S}_n$: swapping any two Hilbert spaces $\h_i, \h_j$ is a symmetry of the mapping $E_n \ra \ket{E_n}$. It also implies that the empty graph is mapped to $\ket{E_n}= \ket{\psi}^{\otimes n}$, with $\ket{\psi}\in \h_1$ a free parameter of the theory. Hence we have:
\begin{proposition}
\label{prop1}
Given a graph $G =(V, E)$ of order $n$, the corresponding quantum state $\ket{G} \in \h$ belongs to a Hilbert space of $n$ identical quantum systems
\be
\h= \h_1^{\otimes n}
\ee
where $\h_1$ is the Hilbert space associated to a single vertex. Moreover, the empty graph is mapped to $E_n \ra \ket{E_n}= \ket{\psi}^{\otimes n}$.
\end{proposition}
The dimension of $\h_1$ is arbitrary and is a free parameter of the theory. This gives us the freedom to consider various families of mappings with finite or infinite Hilbert spaces. Specific examples are $\h_1= \CC^d$ (qudit), $\h_1= \mathrm{span}\{ \ket{k} \}_{k=0}^\infty$ (Fock space of a harmonic oscillator) or the (uncountably) infinite dimensional Hilbert space of a continuos variable.

Axioms {\bf A1} and {\bf A2} determine the structure of the Hilbert space $\h$ and can intuitively be thought of as giving the ``kinematics'' of the model. However, we also need the equivalent of ``dynamics'': given a graph $G$, how do we construct $\ket{G}$? Proposition \ref{prop1} implies that all graphs of order $n$ are mapped to the same Hilbert space $\h= \h_1^{\otimes n}$. Therefore, given two graphs of order $n$, $G_1=(V_1, E_1)$ and $G_2= (V_2, E_2)$, there exists a linear operator $U\in \cal{L}(\h)$ such that
\be
\ket{G_2}= U(G_1, G_2) \ket{G_1}
\ee
In particular, for any graph $G$ we have
\be
\ket{G}= U(G)\ket{E_n}= U(G) \ket{\psi}^{\otimes n}
\label{e_n}
\ee
with the obvious notation $U(G):= U(E_n, G)$. Of course, this only shifts the problem to one of finding $U(G)$. Since the last equation still does not tell us how to find the operator $U(G)$, we supplement our set of axioms with a final one:
\begin{quote}
{\bf A3:} {\bf Universal edge operator.} If the graphs $G= (V, E)$ and $G'= (V', E')$ differ by a single edge, i.e., $V'= V$ and $E'= E \cup \{ (x,y) \}$, then
\be
\ket{G'}= U(x,y) \ket{G}
\label{A3}
\ee
The edge operator $U(x,y)$ is independent of both $G, G'$ and depends only on the edge $(x,y)$.
\end{quote}
Axiom {\bf A3} is particularly strong since it requires the edge operator $U$ to be independent of all graphs. From an axiomatic perspective, it is worth noting that replacing axiom {\bf A3} with a different one (while keeping {\bf A1} and {\bf A2} the same) is analogous to changing the ``dynamics'' of the model, i.e., akin to modifying the ``Hamiltonian" of the system.

Given a graph $G$, axiom {\bf A3} together with eq.~(\ref{e_n}) provides a constructive way to obtain the corresponding quantum state $\ket{G}$: starting from the empty graph $\ket{E_n}$ we apply successively the edge operator corresponding to all graph edges:
\be
\ket{G}= \prod_{(x,y)\in E} U(x,y) \ket{\psi}^{\otimes n}
\label{prod_U}
\ee
Obviously, this construction is consistent only if the edge operator satisfies certain constraints. First, since the graph is undirected, $U(x,y)$ has to be symmetric in its inputs (swapping two vertices is a symmetry of the edge graph $K_2$). Second, the order in which we apply the edge operators in eq.~(\ref{prod_U}) should be irrelevant. Finally, $U$ has a natural local action, as can be seen from the following argument. Consider the graph of order $n$ having a single edge $(x,y)$, $G= (V, \{(x,y)\})$. From {\bf A3} we have $\ket{G}= U(x,y)\ket{E_n}= U(x,y) \ket{\psi}^{\otimes n}$. On the other hand, from the separability axiom {\bf A1} we know that $\ket{G}= \ket{K_2 \uplus E_{n-2}}= \ket{K_2}\otimes \ket{\psi}^{\otimes n-2}$, with $K_2$ an edge. A natural way of satisfying this for all $n$ is to require that the edge operator is local, i.e., acts only on the Hilbert spaces of the corresponding vertices $\h_x\otimes \h_y$. The following three properties summarize the (sufficient) consistency conditions required for the edge operator:

\noindent {\bf C1:} {\em Locality.} The edge operator $U(x,y)$ acts nontrivially only on the Hilbert spaces $\h_x\otimes \h_y$ associated with vertices $x,y$
\be
U(x,y)= U \otimes I^{\otimes n-2}
\label{c1}
\ee
with $I$ the identity acting on the rest. Without the risk of confusion we denote the edge operator by either $U(x,y)$ or $U$.

\noindent {\bf C2:} {\em Symmetry.} For undirected graphs $G$, the edge operator is symmetric in the inputs $U(y,x)= U(x,y)$. Let $S(x,y)= \sum_{i,j} \ket{ij} \bra{ji}$ be the swap operator acting on the Hilbert spaces of two vertices $\h_x\otimes \h_y$. Since $U(y,x):= S(x,y) U(x,y) S(x,y)$, the symmetry condition is:
\be
[U(x,y), S(x,y)]= 0
\label{c2}
\ee

\noindent {\bf C3:} {\em Edge commutativity.} Consider two edges sharing a common vertex. Then the corresponding $U$'s should commute:
\be
[U(x,y), U(y,z)]=0
\label{c3}
\ee
or, equivalently, $[U\otimes I, I \otimes U]=0$.

To summarize, the axiomatic framework presented here defines a class of theories characterized by a triplet
\be
{\cal G}= (\h_1, \ket{\psi}, U)
\label{triplet}
\ee
with $\h_1$ the Hilbert space associated to a vertex, $\ket{\psi}\in \h_1$ the initial state and $U\in {\cal L}(\h_1^{\otimes 2})$ the (local) edge operator. The graph state $\ket{G}$ is constructed from the initial state $\ket{\psi}^{\otimes n}$ by applying the edge operator $U(x,y)$ for each edge $(x,y)\in E$, $\ket{G}= \prod_{(x,y)\in E} U(x,y) \ket{\psi}^{\otimes n}$. This construction is consistent if the edge operator satisfies the conditions {\bf C1}--{\bf C3}. Physically important cases correspond to $U$ a unitary operator, a projector, or a combination of both, as shown in the next section.

\section{Examples}\label{s3}

After developing the general framework in the previous section, we now discuss several important classes of entangled states emerging from the present formalism.

\noindent
(i) {\em Graph states}. ${\cal G}= (\CC^2, \ket{+}, C(Z))$. This is the simplest case and has been studied extensively \cite{graph1,graph2,graph3}. At each vertex there is a qubit ($\h_1= \CC^2$) initialized in $\ket{\psi}= \ket{+}=\frac{1}{\sqrt{2}}(\ket{0}+\ket{1})$. The edge operator is $U=C(Z)= \diag(1,1,1,-1)$. The graph state $\ket{G}$ is constructed by applying a $C(Z)$ operator for each graph edge. An important example are cluster states, an essential resource for the one-way quantum computing model \cite{1wqc}.

\begin{figure}
\putfig{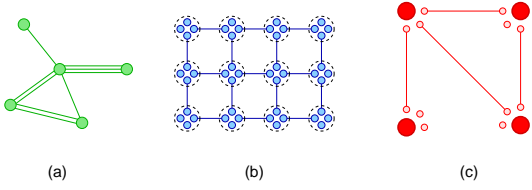}{7}
\caption{(color online). Examples of quantum states emerging from the present formalism: (a) qudit graph states; graph states are obtained for $d=2$; (b) projected entangled pair states (PEPS); (c) quantum random networks \cite{qrn} with $n=4$.}
\label{examples}
\end{figure}

\noindent
(ii) {\em Qudit graph states}. ${\cal G}= (\CC^d, \ket{+}_d, C(Z_d))$, Fig.\ref{examples}(a). These are a straightforward generalization of graph states \cite{qudit_graph}: each vertex is a qudit $\h_1= \CC^d$ initialized in $\ket{+}_d:= \tfrac{1}{\sqrt{d}} \sum_{i=0}^{d-1} \ket{i}$. The edge operator is the controlled-$Z_d$, $U= \sum_{j,k=0}^{d-1} \omega^{jk} \ket{jk}\bra{jk}$, with $Z_d= \diag(1,\omega, \ldots, \omega^{d-1})$ the generalized Pauli-$Z$ and $\omega:= e^{2\pi i/d}$. In this case we can have up to $d-1$ multiple edges between vertices (since $Z_d^d= I$). The construction is still consistent with the axioms {\bf A1}--{\bf A3}, although the graphs are no longer simple. The case $d=2$ corresponds to graph states discussed previously.

\noindent
(iii) {\em Gaussian states and continuous-variable (CV) cluster states}. Gaussian and CV cluster states can be described in a unified framework \cite{CV_graphs}. To each vertex $i$ we associate a harmonic oscillator (qumode) with an infinite dimensional Hilbert space $\h_1= \mathrm{span}\{ \ket{s}_p \}$, where $\ket{s}_p$ are momentum eigenstates, $p\ket{s}_p= s\ket{s}_p$. The initial state $\ket{\psi}$ at each vertex is either an infinitely-squeezed, zero-momentum eigenstate $\ket{0}_p$ (for CV cluster states) or a finitely squeezed state $\ket{\beta}$ (for Gaussian states). For each edge $(i,j)$ we apply a controlled-$Z$ operator $C_{ij}(g)= e^{i g q_i q_j}$, where $q_i$ is the position quadrature for the vertex (mode) $i$ and $g$ the interaction strength (for weighted graphs). The Gaussian/CV cluster state is $\ket{G}= e^{\frac i 2 g\sum_{i,j} \A_{ij} q_i q_j} \ket{\psi}^{\otimes n}$.

The previous three examples share the triplet structure (\ref{triplet}) with a single quantum system (qubit/qudit/qumode) associated to each vertex. However, we are not restricted to a ``monolithic'' Hilbert space -- {\em a priori} we can have an arbitrary quantum system, including a composite one. This opens up new possibilities as shown next. Consider a regular graph of degree $g$ and associate to each vertex $g$ qudits, $\h_1= (\CC^d)^{\otimes g}$. Each qudit subspace $\CC^d$ is paired with the corresponding one from the neighboring vertices, see Fig.\ref{examples}(b) and (c) where the vertex degree is $g=4$ and, respectively, $g=n-1$. The composite structure of $\h_1$ suggests a particular action of the edge operator: $U$ acts only between the corresponding qudits at each vertex, Fig.\ref{examples}(b),(c). Clearly, $U$ is local and obeys trivially the edge commutativity condition {\bf C3}; hence $U$ needs to satisfy only the symmetry constraint {\bf C2} (\ref{c2}).

Moreover, since the mapping $G\ra \ket{G}$ does not have to be invertible, a further generalization is to include projector operators. In this case the edge operator can take the form of a unitary $V$ followed by a projector, $U= \ket{\Psi}\bra{\Psi} V$. However, since the projector does not commute with the unitary part, we can ensure the constraints {\bf C2}--{\bf C3} by applying the projectors after all the unitary operators. The next two examples fit neatly in this class.
 
\noindent
(iv) {\em Projected entangled pair states (PEPS)}. PEPS are higher-dimensional generalization of valence bond states/matrix product states, Fig.\ref{examples}(b). They play an important role in solid-state as ground states of local Hamiltonians -- the AKLT model is a particular example \cite{peps,peps2,peps3}. Each vertex shares with its nearest neighbors a $d$-dimensional maximally entangled state $\ket{\Phi_d}= \tfrac{1}{\sqrt{d}} \sum_{i=0}^{d-1} \ket{ii}= V\ket{+}_d^{\otimes2}$. We construct this state by acting on the initial state $\ket{\psi}= \ket{+}_d^{\otimes2}$ with the symmetric 2-qudit operator $V= d^{-3/2} \sum_{i,j,k,l=0}^{d-1} \omega^{(i-j)(k-l)} \ket{ij}\bra{kl}$. The Hilbert space at each vertex is $\h_1= (\CC^d)^{\otimes g}$, where $g$ is the vertex degree; $g=4$ for the regular 2D lattice in Fig.\ref{examples}(b). PEPS are obtained by projecting the state at each vertex (the $g$ singlets) onto a subspace of dimension $k$ (the ``physical'' subspace). The edge operator $U$ is given by $V$ followed by this projector.

\noindent
(v) {\em Quantum random networks (QRN)}. These states \cite{qrn} are the quantum analog of the classic Erd\H{o}s-R\'enyi random networks. Each graph node shares with the other $n-1$ nodes an entangled state $\ket{\Omega}= \sqrt{1-\frac p 2}\ket{00}+ \sqrt{\frac p 2}\ket{11}$, with $0\le p\le 1$. The Hilbert space at each vertex is $\h_1= (\CC^2)^{\otimes n-1}$, see Fig.\ref{examples}(c). To construct a QRN each pair of nodes tries to convert, using only (stochastic) local operations and classical communication, their shared link into the maximally entangled state $\ket{\Phi^+}= \frac{1}{\sqrt{2}} (\ket{00}+ \ket{11})$. This is equivalent to a projection onto the $\ket{\Phi^+}$ subspace which succeeds with probability $p$. The edge operator is $U= \ket{\Phi^+}\bra{\Phi^+} V$; we choose the symmetric 2-qubit unitary $V$ to satisfy $V\ket{00}= \ket{\Omega}$, giving $V= \sqrt{1-\frac p 2} I^{\otimes 2}+ \frac i 2 \sqrt{\frac p 2}(\sigma_x- \sigma_y)^{\otimes 2}$; the initial state is $\ket{\psi}= \ket{0}^{\otimes n-1}$.

So far we discussed well-known examples which fit in the above axiomatic framework. Are there any novel quantum states which emerge from this framework? In the following we present the general solution for qubits, $d=2$, which contains (and generalizes) graph states. Imposing the constraints {\bf C1}--{\bf C3} we obtain two families of solutions for the edge operator $U$. The first is diagonal:
\be
U_I= \diag(a,b,b,c)
\label{U1}
\ee
Two important examples belong to this family:\\
(a) {\em Graphs states}. Taking $a=b=-c=1$ we recover the construction for graphs states given above. More generally, the edge operator is a product of single-qubit phase shifts and a controlled phase shift, $U= \diag(1,1,1,e^{i\varphi}) P(\alpha)^{\otimes 2}$, with $P(\alpha):= \diag (1,e^{i\alpha})$.

~\\
(b) {\em Parity projectors}. The two parity operators $P_0= \diag(1,0,0,1)$ and $P_1= \diag(0,1,1,0)$ project, respectively, on the even parity, $\mathrm{span}\{ \ket{00}, \ket{11}\}$, and odd parity, $\mathrm{span}\{ \ket{01}, \ket{10}\}$ subspaces of two qubits. Parity gates, together with single-qubit gates, are universal for quantum computation \cite{spin_parity,ri_parity}.

The second solution is parametrized by $a, b, c$ and $T$:
\ba
U_{II}= a I^{\otimes 2} + b(T\otimes I+ I\otimes T) + c T\otimes T \label{U}
\label{U2}
\ea
with $T= \begin{bmatrix}0 & 1 \cr \gamma & -\alpha \end{bmatrix}$ or $T= \begin{bmatrix}0 & \gamma \cr 1 & -\alpha \end{bmatrix}$.

There are some intriguing similarities between the present approach and quantum lattice gas models \cite{qlg,qlg2} which can prove insightful for future research.

\section{Extension}\label{s4}

The modular structure of our framework (given by the axioms {\bf A1}--{\bf A3} and consistency conditions {\bf C1}--{\bf C3}) enables us to generalize it to non-simple graphs. In this section we extend the formalism to directed graphs and weighted graphs. For directed graphs, we replace the consistency conditions {\bf C2}--{\bf C3} while keeping the others unchanged. For weighted graphs, the edge operator will be different for different edges.

\noindent
{\em Directed graphs}. Since in this case the edge operator $V$ is no longer symmetric, $V(y,x) \ne V(x,y)$, we need to change the consistency conditions {\bf C2}--{\bf C3} (locality {\bf C1} still holds). Let $V'(x,y)$ be the swapped version of the edge operator
\be
V'(x,y):= V(y,x)= S(x,y) V(x,y) S(x,y)
\ee
We represent the two operators graphically as quantum networks:
\[
V(x,y):= \insertfig{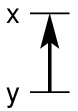}{0.65} \ \ \ \mathrm{and}\ \ \  V'(x,y):= \insertfig{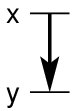}{0.65}.
\]
The previous commutations relations on two and three vertices now become:
\be
\mathrm{\bf D2:}\ \ \  [V(x,y), V'(x,y)]=0, \ \ \ \insertfig{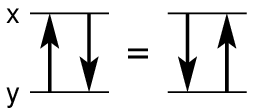}{1.9}
\label{d2}
\ee
and
\be
\mathrm{\bf D3a:}\ \ \  [V(x,y), V(y,z)]= 0, \ \ \ \insertfig{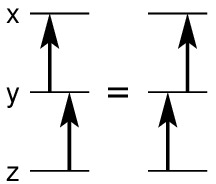}{1.7}
\ee
\be
\mathrm{\bf D3b:}\ \ \  [V(x,y), V'(y,z)]= 0, \ \ \ \insertfig{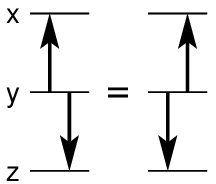}{1.7}
\ee
\be
\mathrm{\bf D3c:}\ \ \  [V'(x,y), V(y,z)]= 0, \ \ \ \insertfig{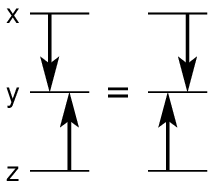}{1.7}
\ee
The fourth condition $[V'(x,y), V'(y,z)]=0$ is equivalent to {\bf D3a}: $0= S(x,z) [V(x,y), V(y,z)] S(x,z)= [V(z,y), V(y,x)]= [V'(y,z), V'(x,y)]$.

Any solution of the edge operator for directed graphs automatically provides a solution for undirected graphs. Let $V(x,y)$ be a directed edge operator satisfying the constraints {\bf D2}--{\bf D3}. Define
\[
U(x,y):= V(x,y) V'(x,y)
\]
Using the graphical notations introduced above, it is straightforward to prove that $U(x,y)$ satisfies the constraints {\bf C2}--{\bf C3} and thus $U$ is an edge operator for undirected graphs. This solution for undirected graphs captures intuitively the well-known fact from graph theory that an undirected edge is equivalent to two oppositely-directed edges (as can be seen from the entries of the adjacency matrix of the graph).

An example of a unitary solution for directed edge operator for qubits ($d=2$) is given pictorially by
\[
\insertfig{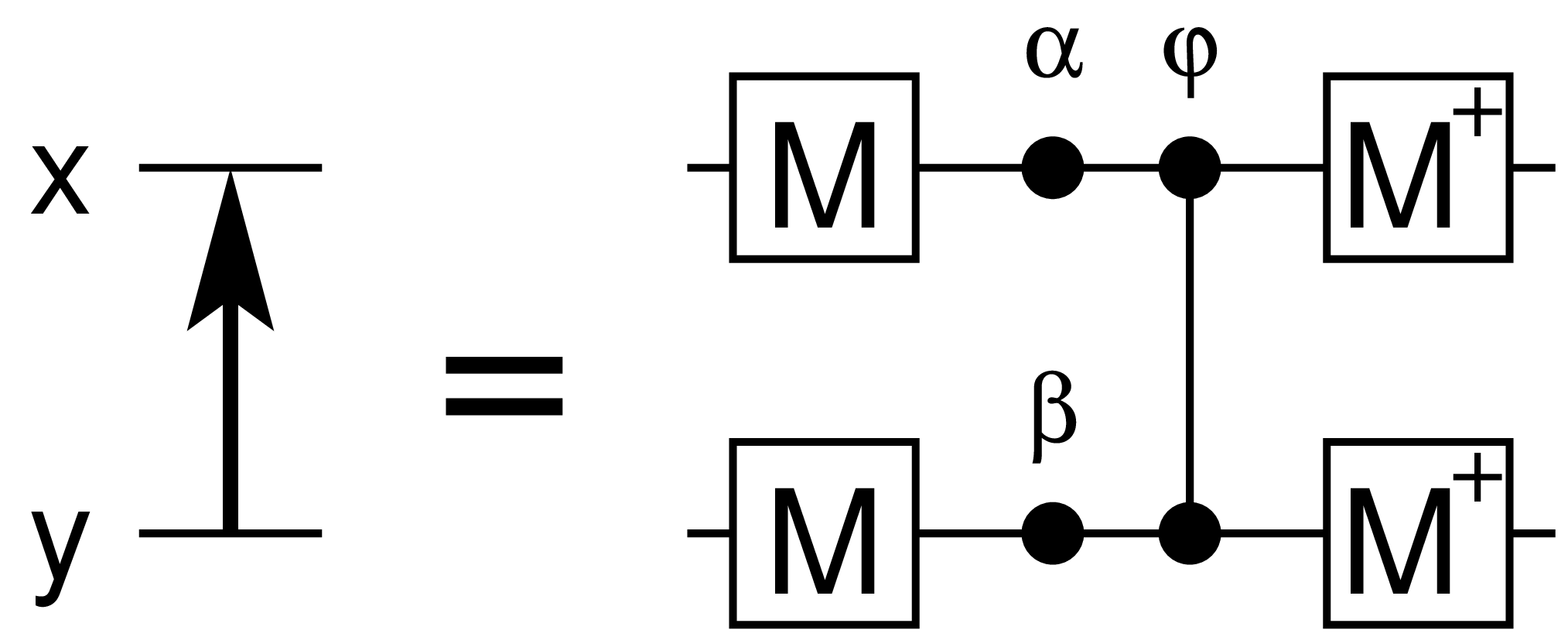}{3},
\]
with $M$ an arbitrary unitary, $\alpha, \beta$ and $\varphi$ free parameters; equivalently
\be
V(x,y)= M^{\dag\otimes 2} \diag(1,1,1,e^{i\varphi}) P(\alpha)\otimes P(\beta) M^{\otimes 2}
\label{Vdirected}
\ee
In this case the edge asymmetry condition is enforced by choosing $\alpha > \beta$.

\noindent
{\em Weighted graphs}. Several important problems in graph theory involve weighted graphs, in which the edges are labeled by one (or more) parameters (the weight). Notable examples are the traveling salesman problem (the edge weight is the distance between the nodes) or problems in network optimization (weights are the network capacity of the link). In general a weight can have several numbers, each characterizing a different parameter of the link.

The solutions for the edge operator (in both undirected and directed cases) contain free parameters, see eqs.~(\ref{U1}), (\ref{U2}), (\ref{Vdirected}). This suggests a straightforward generalization of the present formalism to weighted graphs. We construct a quantum state associated to a weighted graph by choosing the parameters for the edge operator to be different for different edges (provided they satisfy the constraints {\bf C2}--{\bf C3} or {\bf D2}--{\bf D3}).

Since the solution (\ref{U1}) is diagonal, the parameters $a, b, c$ can be taken different for all edges. Modulo an overall phase, this gives two free parameters to label an edge. For the solutions (\ref{U2}) and (\ref{Vdirected}), the matrices $T$, and respectively $M$, are fixed for all edges in order to ensure edge commutativity ({\bf C3} or {\bf D3}). In this case only the parameters $a,b,c$, and respectively $\alpha,\beta,\varphi$, are free to label different edges.

For undirected graphs, eq.~(\ref{U1}) corresponds to weighted graph states which arise naturally from an Ising-type interaction between spins located on a graph \cite{hartmann,xue}. The solution (\ref{Vdirected}) is a straightforward generalization to directed graphs, where the asymmetry of the edge operator appears from an extra local phase.

\section{Conclusion}\label{s5}

Graphs play an important role in characterizing efficiently several families of multipartite quantum states. A notable example are cluster states which are an essential resource for the one-way quantum computing paradigm. Due to the visual impact of a graph, it is easier to understand the entanglement content of the associated graph state. For instance, the entropic area law \cite{srednicki, hiz12, plenio, hiz3, area_law} can be easily understood in this picture: the entanglement entropy for a bipartition $(A,B)$ of a lattice spin system is proportional to the number of links crossing the boundary between $A$ and $B$.

In this article we have developed an axiomatic framework for mapping graphs to states of a quantum system. Starting from three general axioms we derived a rich structure which includes and generalizes several classes of multipartite entangled state, like graph/stabilizer states, quantum random networks and PEPS. Due to its modular structure (in terms of axioms and consistency conditions), the axiomatic approach developed here is remarkably flexible. By changing some of the consistency conditions while keeping the others the same, we can incorporate in the model non-simple graphs, namely directed and weighted graphs. Specifically, directed graphs can be included by modifying the consistency conditions {\bf C2}--{\bf C3} in order to take into account the edge asymmetry. 

There are several directions in which the present research can be developed in the future. First, one would like to find the general solution of the edge operator (for both undirected and directed graphs) in the case of a qudit, $\h_1= \CC^d$. Second, given a graph $G$ and an edge operator $U$, we would like to characterize the entanglement of the resulting state $\ket{G}$ as a function of the entangling power \cite{entpower} of the edge operator. Third, one can envisage a more general extension by changing the ``dynamics'' of the system by modifying Axiom {\bf A3}. The tensor product structure of $\h$ remains the same (following from {\bf A1}--{\bf A2}), but the edge operator and the consistency conditions will be different. Last but not least, it will be valuable to apply methods based on this approach to graph-theoretical problems (e.g., finding novel graph invariants).

We acknowledge financial support from EU project QAP and from ISI Foundation, Torino, Italy. R.I.~is grateful to Paolo Zanardi and Dirk Schlingemann for illuminating discussions.


\end{document}